\documentclass[twoside]{ilcws08}
\usepackage[latin1]{inputenc}
\usepackage[dvips]{graphicx,epsfig,color}
\usepackage{wrapfig,rotating}
\usepackage{amssymb,amsmath,array}

\def\pbarp{$p\overline{p}$}             
\def\ttbar{$t\overline{t}$}             
\def\etal{{\sl et al.}}                 
\def\ifb{fb$^{-1}$}

\begin{document}
\title{
Top Quark Physics at the Tevatron} 
\author{Cecilia E. Gerber
\thanks{on behalf of the CDF and the D0 collaborations}
\vspace{.3cm}\\
University of Illinois at Chicago \\
845 W Taylor St. M/C 273, Chicago, IL, 60607
}

\maketitle

\begin{abstract}
I present recent results on top quark production and properties 
in \pbarp\ collisions at a center of mass energy of
$1.96\;\rm TeV$. The measurements were performed by 
the CDF and D0 collaborations using approximately 3 \ifb of data
taken during Run II at the Tevatron.
\end{abstract}

\section{Introduction}

The top quark was
discovered at the Fermilab Tevatron Collider
in 1995~\cite{topdiscoveryI,topdiscoveryII}
and
completes the quark sector of the three-generation structure of the
standard model (SM). It is the heaviest known elementary particle with a 
mass approximately
40 times larger than that of the next heaviest quark, the bottom quark. 
It differs from the
other quarks not only by its much larger mass, but also by its
lifetime which is too short to build hadronic bound states.
The top quark is one of the least-studied components of the SM,
and the Tevatron, with a center of mass energy 
of $\sqrt{s} = 1.96\;\rm TeV$, is at present
the only accelerator where it has been produced.
The top quark plays an important role in the
discovery of new particles, as the Higgs boson
coupling to the top quark is stronger than to all other
fermions. Understanding the production properties of
top quark pairs is in itself a test perturbative 
Quantum Chromo Dynamics (pQCD), but is also a 
crucial ingredient in the discovery
of new physics beyond the SM. 
In the following sections I will present studies of top quark production
and decay mechanisms, both within and out of the SM predictions. 

\section{Studies of \ttbar\ pair production mechanisms}
At Tevatron energies, top quarks are produced predominantly in pairs. 
Within the SM, the top quark decays almost exclusively
into a $W$ boson and a $b$ quark, resulting in two $W$ bosons and 2 
$b$ jets in each \ttbar\ pair event. 
The $W$ boson itself decays into one lepton and its 
associated neutrino, or hadronically. 
We have classified the \ttbar\ pair decay channels as follows:
the dilepton channels where both $W$ bosons decay leptonically 
into an electron or a muon ($ee$, $\mu\mu$, $e\mu$), 
the lepton + jets channels where one of the $W$
bosons decays leptonically and the other hadronically ($e$+jets,
$\mu$+jets), and the all-jets channel where both $W$ bosons decay
hadronically. Production cross sections have been measured in all
decay channels. The lepton + jets channels have less statistics than the 
all-jets channel, but the background level is significantly smaller, making
it the channel of choice for the measurement of top quark properties. 

The total top quark pair production cross section for a hard
scattering process initiated by a $p\overline{p}$ collision 
is a function of the top quark mass
$m_t$. 
For a top quark mass of $175\;\rm GeV$, the predicted
SM \ttbar production cross section is 
$6.7^{+0.7}_{-0.9}\;\rm pb$~\cite{theoxsec}.
Deviations of the measured cross section from the theoretical
prediction could indicate effects beyond QCD perturbation theory.
Explanations might include substantial non-perturbative effects, new
production mechanisms, or additional top quark decay modes beyond the SM.
Previous measurements~\cite{CDFRun2,D0Run2} show 
good agreement with the theoretical expectation within the experimental 
precision. 

\begin{figure*}[htb]
\centerline{\hbox{{\epsfysize=8.5cm 
\epsffile{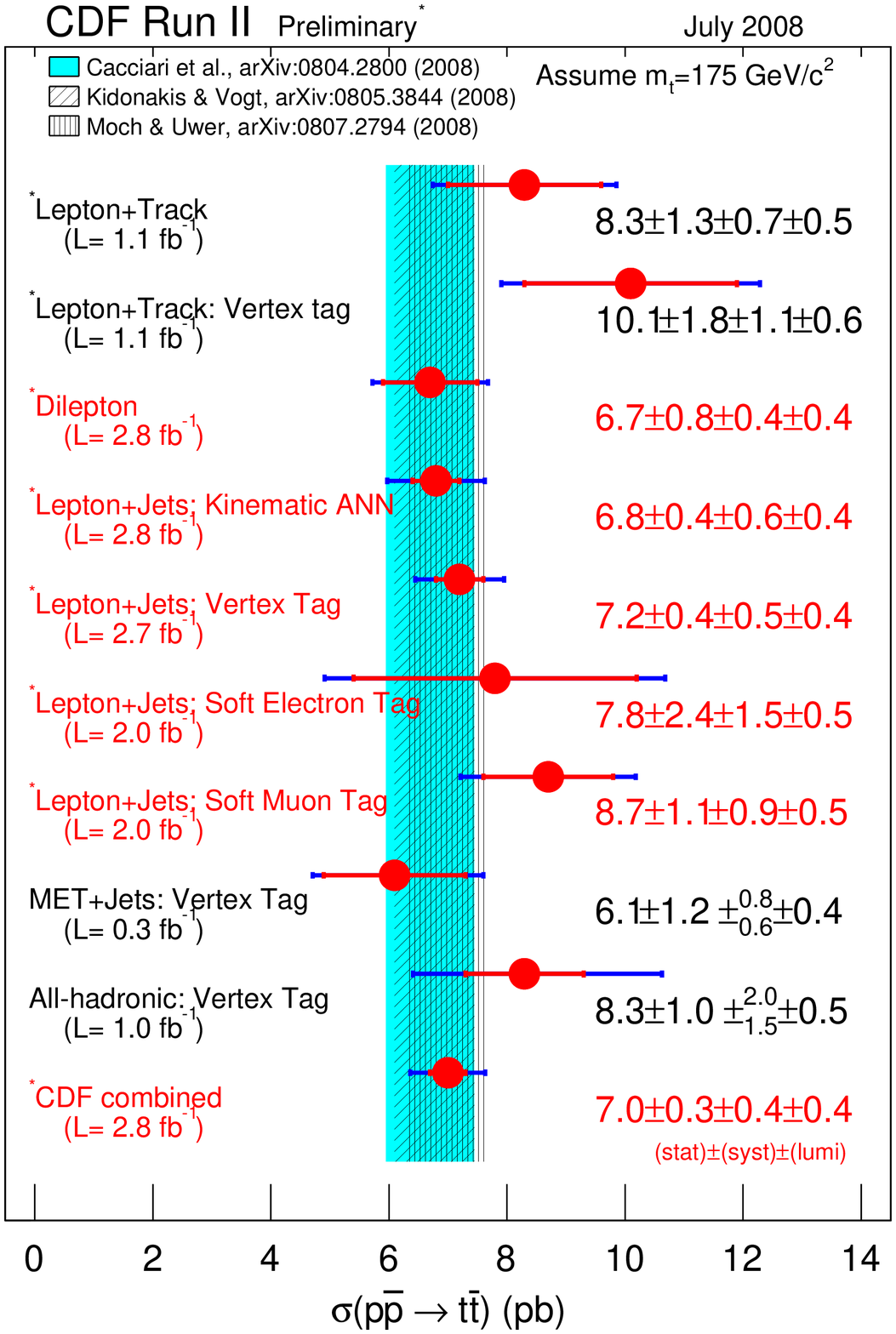}}
{\epsfysize=8.5cm \epsffile{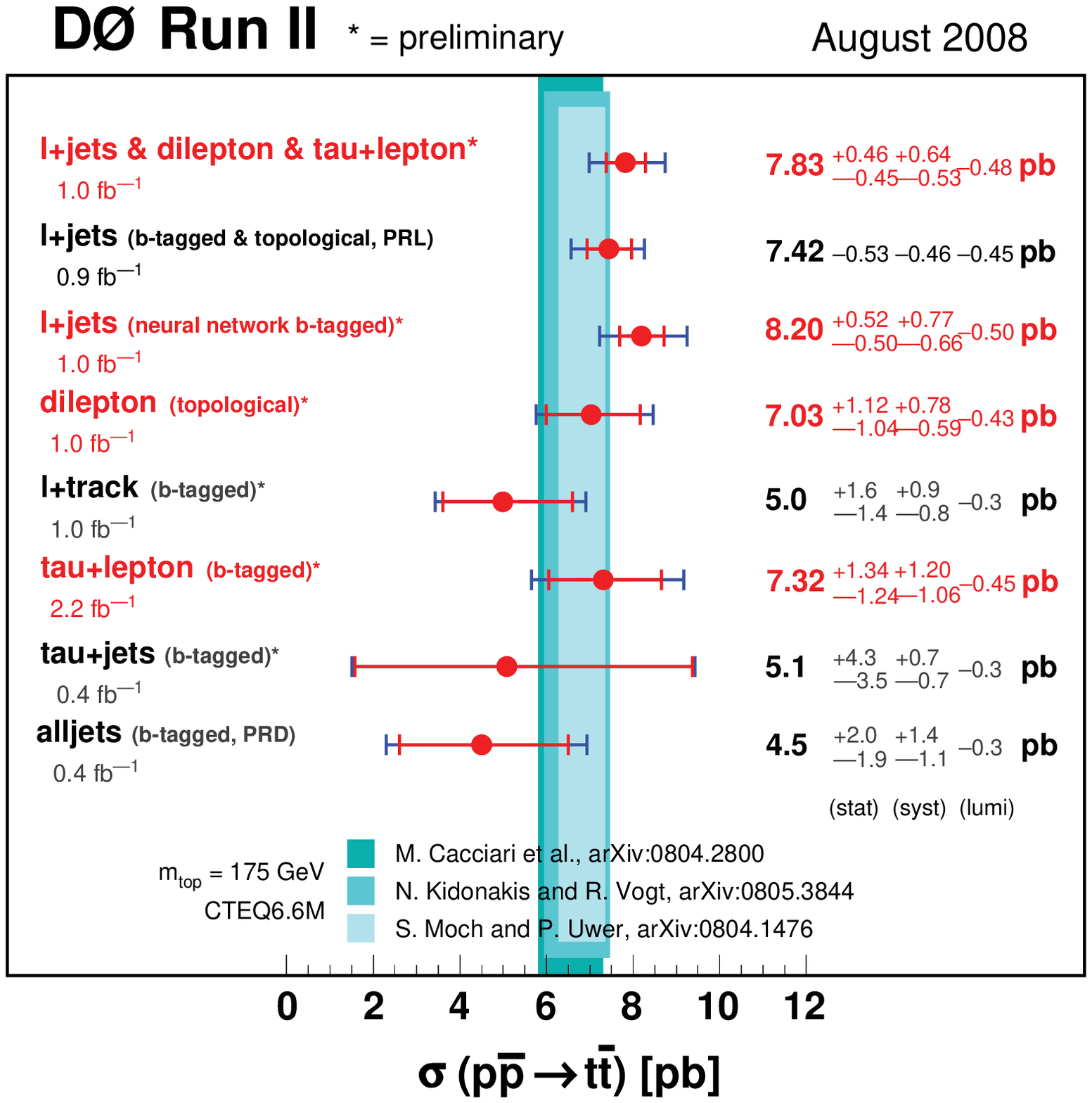}}}}
\caption{Summary of \ttbar pair production cross section
measurements at $\sqrt{s} = 1.96\;\rm TeV$ from CDF (left), and D0 (right). 
Theoretical predictions are shown as vertical bands. The newest experimental
results have reached the theoretical uncertainty of $\approx$~10\%. 
}
\label{fig:xsec}
\end{figure*}

Recent results are 
available from both CDF and D0, and are summarized in 
Fig.~\ref{fig:xsec}, together with the theoretical predictions. 
As can be seen in the plots, the uncertainties on the 
latest experimental results are reaching the theoretical uncertainty of
$\approx$~10\%.  

At leading order, \ttbar\ 
production proceeds through the $q\bar q \rightarrow t\bar t$ and $gg
\rightarrow t\bar t$ processes, with the $q\bar q$ process contributing
85\% to the production cross section, and  the $gg$ 
process contributing only 15\%. NLO theoretical predictions are 
available~\cite{theoxsec}, but suffer from 
large uncertainties. Measuring the relative fraction of \ttbar\ events 
produced via a particular production mechanism 
provides a direct test of pQCD and 
may reveal the existence of \ttbar\ production and
decay mechanisms beyond the ones predicted by the SM~\cite{kane}. 
CDF has studied the 
relative fraction of \ttbar\ events produced via gluon-fusion 
$\frac{\sigma(gg\rightarrow
t\overline{t})}{\sigma(q\overline{q}\rightarrow t\overline{t})}$ 
using an azimuthal correlation of charged leptons in the 
\ttbar\ to dilepton channel.
Using 2 \ifb of data, CDF 
measures 
$\frac{\sigma(gg\rightarrow
t\overline{t})}{\sigma(q\overline{q}\rightarrow
t\overline{t})}=0.53\pm^{0.36}_{0.38}$.
The measurement is dominated by the statistical uncertainty and is 
in agreement with SM expectations.


Several beyond the SM theories~\cite{resonance} predict the resonant 
production of \ttbar\ pairs. 
Using 2.1~\ifb of data, D0 has studied the
invariant mass spectrum in lepton + jets events
using a neural network to identify b-jets. 
The observed spectrum is consistent with 
SM expectations, showing no evidence for additional resonant production 
mechanisms. Consequently, the data is used to 
set upper limits on $\sigma \times B(X \rightarrow t\overline{t})$ 
for different hypothesized 
resonance masses using a Bayesian
approach. Within a topcolor-assisted technicolor model, 
the existence of a leptophobic Z boson
with mass $m_{Z\prime} < 760\;\rm GeV$ and width 
$\Gamma_{Z\prime} = 0.012 m_{Z}$ 
can be excluded at 95\% C.L.

\section{Studies of top quark decays}

Within the SM, the top quark decays via the V-A
charged-current interaction to a $W$ boson and a $b$ quark.
New physics present in the $t \to Wb$ decay could become evident in the 
helicity of the $W$ boson originating from a top quark decay. 
In particular, a different Lorentz structure of the $Wtb$ interaction
would alter the fractions of $W$ bosons produced in each polarization state 
from the SM values of $0.697\pm0.012$~\cite{WHel1} for the 
longitudinal $f^0$ 
and
$3.6 \times 10^{-4}$~\cite{WHel2} 
for the right-handed $f^+$ polarization.  
The CDF and D0 collaborations have measured these fractions using as sensitive
observable the cosine of the decay angle of the charged lepton in the $W$ rest 
frame measured with respect to the direction of motion of the $W$ boson in the 
top quark rest-frame. The measured fractions using $1.9\;\rm fb^{-1}$ of CDF
data~\cite{WHel3} and $2.7\;\rm fb^{-1}$ of D0 data~\cite{WHel4} 
are limited by 
the statistics of the data samples and show no significant deviation from the 
SM predictions.

\section{Measurement of the top quark mass}

The mass of the top quark is a fundamental parameter of the SM. When
combined with the $W$ mass measurement, it places constraints on the mass of the
Higgs boson. Both collaborations have measured the top quark mass in different
decay channels and with different methods. The most precise results are obtained
in the lepton + jets channel using the Matrix Element technique~\cite{ME}. In
this method, an event by event weight is calculated according to the quality 
of the agreement with the SM top quark pair and background 
differential cross-sections; the product of all event probabilities gives the 
most likely mass. The jet energy scale is constrained in-situ 
by the hadronic decay of the $W$ boson present in the event. A summary of all the
measurements is shown in figure~\ref{fig:mass}. The combined 
result~\cite{world} of 
$m{\rm(top)}=172.4\pm0.7{\rm(stat)}\pm1.0{\rm(syst)}\;\rm GeV$ has a precision
of less than 1\%. 

Assuming that the top pair production is governed by the SM, the D0
collaboration extracted the top quark mass comparing the measured cross-section 
with a theoretical prediction~\cite{moch}, and obtained 
$m{\rm(top)}=169.6+5.4-5.5\;\rm GeV$~\cite{d0mass}. This 
measurement has different experimental and theoretical uncertainties than the 
direct measurements; the results obtained by the two methods are in agreement
with each other.

\begin{figure*}[htb]
\centerline{\hbox{{\epsfysize=8.5cm 
\epsffile{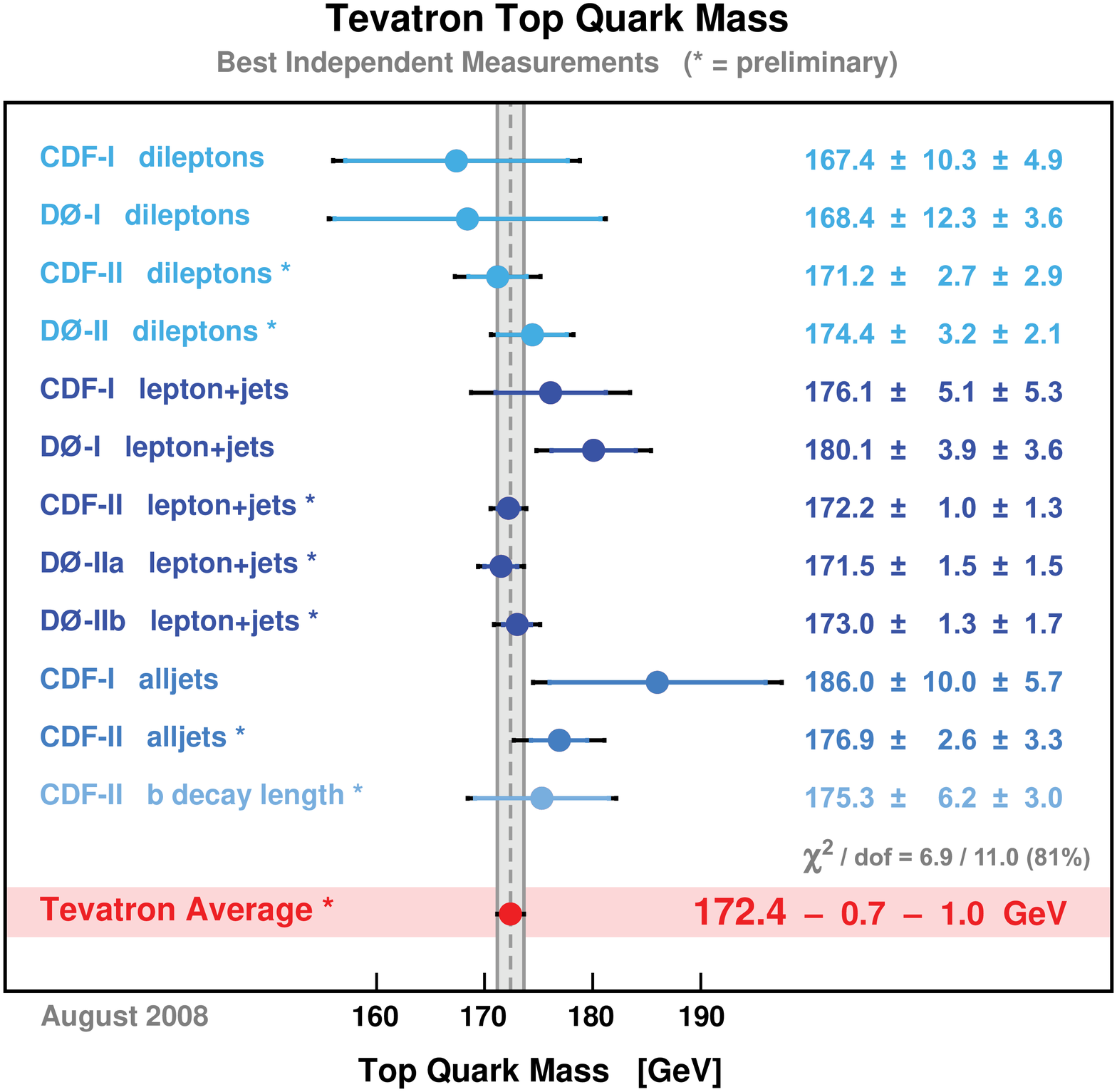}}
{\epsfysize=8.5cm \epsffile{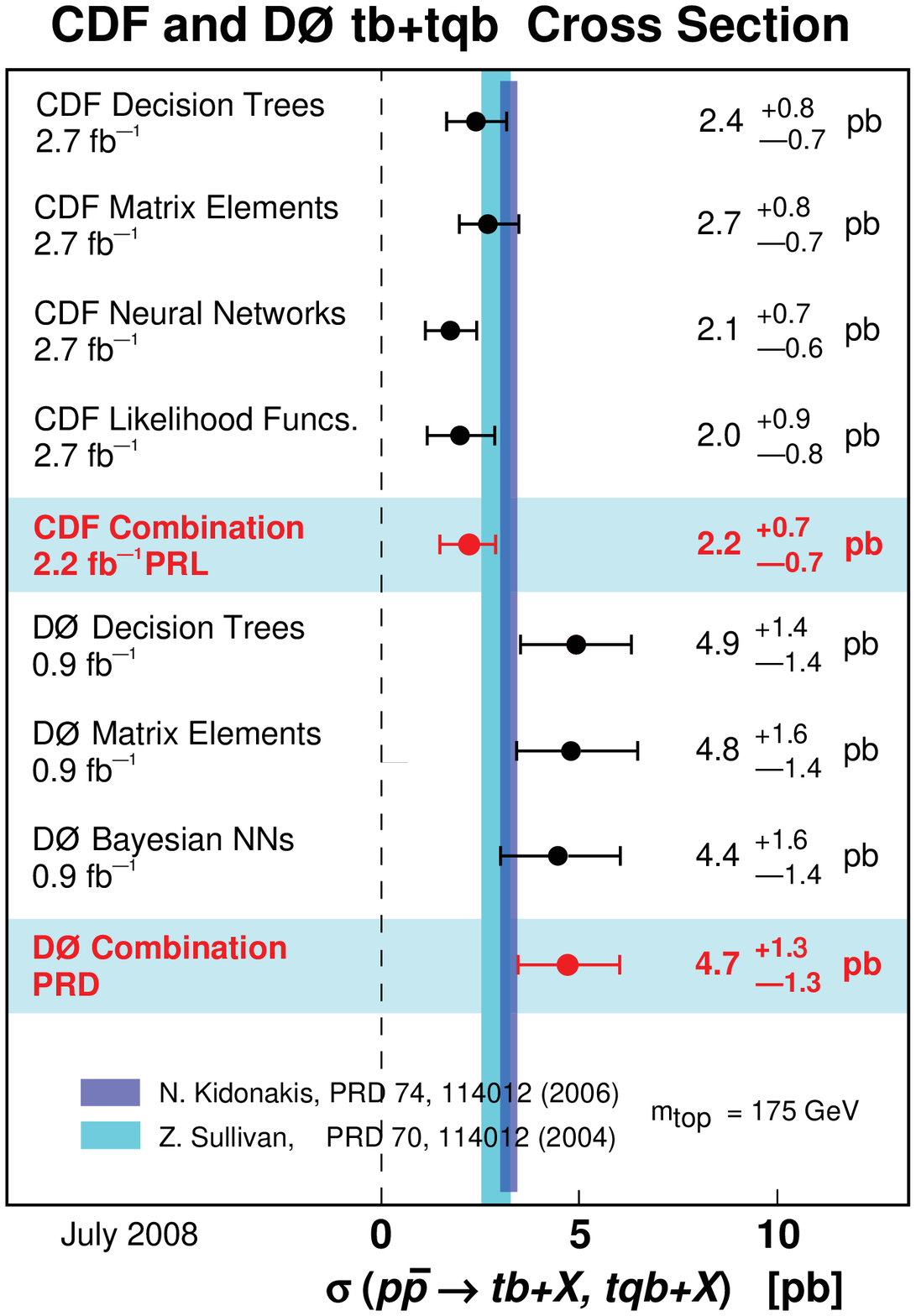}}}}
\caption{Left: Summary of the top quark mass 
measurements and the world average. Right: Summary of the single top production cross section 
measurements at 
$\sqrt{s}=1.96\;\rm TeV$.  
Theoretical predictions are shown as vertical bands.}
\label{fig:mass}
\end{figure*}

\section{Single top production}

The SM predicts that top quarks can be produced singly at hadron 
colliders~\cite{singletop-willenbrock} 
via the electroweak interaction from the decay  
of an off-shell $W$~boson or fusion of a virtual $W$~boson
with a $b$~quark.
The SM prediction~\cite{singletop-xsec} 
for the single top quark cross section for a top quark 
mass of $175\;\rm GeV$ is $0.98\pm0.04\;\rm pb$
for the s-channel process $p\overline{p} \to tb$, and 
$2.16\pm0.12\;\rm pb$ for the t-channel process 
$p\overline{p} \to tqb$. 
Measurement of the
single top quark production cross section has been impeded by its low
rate and difficult background environment compared to the top pair production. 
The D0 collaboration presented in 
2007 the first evidence for single top quark production
and the first direct measurement of
$|V_{tb}|$~\cite{d0-prl-2007,d0-prd-2008} based on 0.9~fb$^{-1}$ of
data. More recently, the
CDF collaboration has also presented evidence for single top quark
production~\cite{cdf-prl-2008} in 2.2~fb$^{-1}$ of data. The summary of these
results is shown in figure~\ref{fig:mass}. 

Events with single top quarks have also been used by both collaborations 
to directly measure the
absolute value of the CKM matrix element $|V_{tb}|$, and to search for physics
behond the SM. In particular, the D0 collaboration searched for a charged Higgs boson $H^+$
with $180<m_H^+<300\;\rm GeV$ decaying to a top and a bottom quark, which 
would enhance the rate for the s-channel single top production and be observed
as a
resonance in the $m(Wjj)$ spectrum~\cite{H+}; the D0 collaboration also used the
data to place constraints on the 
left-handed and right-handed
vector and tensor $Wtb$ couplings~\cite{Wtb}. 
The CDF collaboration performed a model independent search for FCNC processes in
single top production, and set an upper limit on the production cross section 
$\sigma(u(c)+g\to t)<1.8\;\rm pb$ at the 95\% C.L. limit and on the anomalous
couplings that define the strength of the $gtu$ and the $tcg$ 
vertices~\cite{FCNC}. 

\section{Conclusions}

The Tevatron has entered a new era of top quark precision measurements. 
The experimental
precision on the top quark pair production cross section results has reached the 
theoretical uncertainty, making comparisons between different channels 
and methods interesting. In
addition, a series of new measurements of top quark production and decay 
properties and searches for deviations from the SM predictions are 
becoming available based
on the larger statistics samples collected in 3~\ifb of collider data. 
CDF and D0 have already written to tape more than twice the amount of data 
used for
these results. The lepton $+ \ge 3$ jets sample with two identifed $b$-jets is 
completely dominated by \ttbar\ events. With larger data sets, 
as the ones that will be available at the Tevatron in the
near future, even more precise measurements 
of top quark properties will be possible.
 
\section{Acknowledgments}

I would like to thank my collaborators from the CDF and D0 collaborations 
for their help in preparing this document.
I also thank the staffs at Fermilab and collaborating institutions, 
and acknowledge support from the 
National Science Foundation (USA). 


\begin{footnotesize}




\begin{thebibliography}{99}
\bibitem{url} Presentation: \\ 
\verb$http://ilcagenda.linearcollider.org/getFile.py$ \\
\verb$/access?contribId=94&sessionId=18&resId=0&materialId=slides&confId=2628$
\bibitem{topdiscoveryI}
F. Abe {\etal}, CDF Collaboration, Phys. Rev. Lett. {\bf 74} 2626 (1995).

\bibitem{topdiscoveryII}
S. Abachi {\etal}, D0 Collaboration, Phys. Rev. Lett. {\bf 74} 2632 (1995).




\bibitem{theoxsec}
N. Kidonakis and R. Vogt,
Phys. Rev. D {\bf 68}, 114014 (2003). 
M.~Cacciari, S. Frixione, M.~L.~Mangano, P.~Nason, and G.~Ridolfi,
JHEP {\bf 0404}, 68 (2004).


\bibitem{CDFRun2}
D. Acosta {\etal}, CDF Collaboration,
Phys. Rev. D {\bf 74}, 072005 (2006);
Phys. Rev. Lett. {\bf 96}, 202002 (2006);
Phys. Rev. Lett. {\bf 97}, 082004 (2006);
Phys. Rev. Lett. {\bf 100}, 072005 (2007);
Phys. Rev. D {\bf 76}, 072009 (2007).
\bibitem{D0Run2}
V. M. Abazov {\etal}, D0 Collaboration, Phys. Lett. B {\bf 626}, 35 (2005); 
Phys. Rev. D {\bf 74}, 112004 (2006);
Phys. Rev. D {\bf 76}, 052006 (2007);
Phys. Rev. D {\bf 76}, 072007 (2007);
Phys. Rev. D {\bf 76}, 092007 (2007);
Phys. Rev. Lett. {\bf 100}, 192003 (2008);
Phys. Rev. Lett. {\bf 100}, 192004 (2008).
\bibitem{kane}
G.L. Kane and S. Mrenna, Phys. Rev. Lett. {\bf 77}, 3502-3505 (1996).

\bibitem{resonance}
CDF Collaboration public result:\\
http://www-cdf.fnal.gov/physics/new/top/2006/mass/mttb/mttb\_pub.pdf

\bibitem{WHel1}
G. L. Kane, G. A. Ladinsky, and C.-P. Yuan, Phys. Rev. D {\bf 45}, 124 (1992); 
R. H. Dalitz and G. R. Goldstein, ibid., 1531;
C. A. Nelson et al., Phys. Rev. D {\bf 56}, 5928 (1997).

\bibitem{WHel2}
M. Fischer et al., Phys. Rev. D {\bf 63}, 031501(R) (2001).

\bibitem{WHel3}
CDF Collaboration public result:\\
http://www-cdf.fnal.gov/physics/new/top/summaryplots/whel.eps

\bibitem{WHel4}
D0 Collaboration public result:\\
http://www-d0.fnal.gov/Run2Physics/WWW/results/prelim/TOP/T69/T69.pdf

\bibitem{ME}
V. M. Abazov {\etal}, D0 Collaboration, 
Nature {429}, 638 {2004}. 

\bibitem{world}
The Tevatron Electroweak Working Group, 
arXiv:0808.1089 [hep-ex] (2008).

\bibitem{moch}
Sven Moch and Peter Uwer, Phys.\ Rev.\ D {\bf 78}, 034003 (2008). 

\bibitem{d0mass}
D0 Collaboration public result:\\
http://www-d0.fnal.gov/Run2Physics/WWW/results/prelim/TOP/T72/T72.pdf

\bibitem{singletop-willenbrock}
S.S.D.~Willenbrock and D.A.~Dicus,
Phys.\ Rev.\ D {\bf 34}, 155 (1986);

\bibitem{singletop-xsec}
Nikolaos Kidonakis, 
Phys.\ Rev.\ D~{\bf 74}, 114012 (2006).

\bibitem{d0-prl-2007}
V.M.~Abazov {\it et al.} (D0 Collaboration)
Phys.\ Rev.\ Lett.\ {\bf 98}, 181802 (2007).


\bibitem{d0-prd-2008}
V.M.~Abazov {\it et al.} (D0 Collaboration)
Phys.\ Rev.\ D\ {\bf 78}, 012005 (2008).

\bibitem{cdf-prl-2008}
T.~Aaltonen {\it et al.} (CDF Collaboration)
Phys.\ Rev.\ Lett.\ {\bf 101}, 252001 (2008).

\bibitem{H+}
V.M.~Abazov {\it et al.} (D0 Collaboration)
Phys.\ Rev.\ Lett.\ {\bf 102}, 191802 (2009).

\bibitem{Wtb}
V.M.~Abazov {\it et al.} (D0 Collaboration)
Phys.\ Rev.\ Lett.\ {\bf 102}, 092992 (2009).

\bibitem{FCNC}
T.~Aaltonen {\it et al.} (CDF Collaboration)
ArXiv:0812.3400v2, [hep-ex] (2008).

\end{thebibliography}
%

\end{footnotesize}

\end{document}